\documentclass{article}

\PassOptionsToPackage{numbers, compress}{natbib}

\usepackage[final]{dlde_neurips_2023}
\usepackage{amssymb}
\usepackage{amsmath}
\usepackage{multirow}
\usepackage{graphicx}
\usepackage{wrapfig}
\usepackage{booktabs}       
\usepackage{pifont}

\usepackage[utf8]{inputenc} 
\usepackage[T1]{fontenc}    
\usepackage{hyperref}       
\usepackage{url}            
\usepackage{booktabs}       
\usepackage{amsfonts}       
\usepackage{nicefrac}       
\usepackage{microtype}      
\usepackage{xcolor}         

\title{PINNs-TF2: Fast and User-Friendly Physics-Informed Neural Networks in TensorFlow V2}

\author{
  Reza Akbarian Bafghi\\
  University of Colorado, Boulder\\
\texttt{reza.akbarianbafghi@colorado.edu} \\
  \And
  Maziar Raissi \\
  University of California, Riverside\\
\texttt{maziar.raissi@ucr.edu} \\
}

\begin{document}


\maketitle

\begin{abstract}
Physics-informed neural networks (PINNs) have gained prominence for their capability to tackle supervised learning tasks that conform to physical laws, notably nonlinear partial differential equations (PDEs). This paper presents "PINNs-TF2", a Python package built on the TensorFlow V2 framework. It not only accelerates PINNs implementation but also simplifies user interactions by abstracting complex PDE challenges. We underscore the pivotal role of compilers in PINNs, highlighting their ability to boost performance by up to 119x. Across eight diverse examples, our package, integrated with XLA compilers, demonstrated its flexibility and achieved an average speed-up of 18.12 times over TensorFlow V1. Moreover, a real-world case study is implemented to underscore the compilers' potential to handle many trainable parameters and large batch sizes. For community engagement and future enhancements, our package's source code is openly available at: \href{https://github.com/rezaakb/pinns-tf2}{\texttt{https://github.com/rezaakb/pinns-tf2}}.
\end{abstract}

\section{Introduction}
Physics-informed neural networks (PINNs) are gaining traction as a potent tool for supervised learning, ensuring solutions align with physics laws, notably nonlinear partial differential equations (PDEs) \cite{Raissi2019PhysicsinformedNN}. Their versatility covers an array of applications \cite{Haghighat2021APD,Mao2020PhysicsinformedNN,Jeong2023ACP,RashtBehesht2021PhysicsInformedNN,Chen2019PhysicsinformedNN}. This paper unveils ``PINNs-TF2'', a novel Python package to bolster PINNs' efficiency and to simplify the integration of machine learning and physical sciences by abstracting complex PDE problems.

We selected TensorFlow V2 (TF2) due to its prominence in the deep learning realm and its provision of static computational graphs. Given that PINNs frequently necessitate multiple gradient computations of network outputs in relation to inputs for PDE definition \cite{Paszke2017AutomaticDI}, the advantage of static graphs becomes evident. They reduce the overhead that can be significantly time-consuming in dynamic computational graphs, as seen in frameworks like PyTorch \cite{Paszke2019PyTorchAI,Paszke2017AutomaticDI}.

Building upon previous works \cite{McClenny2021TensorDiffEqSM,Hennigh2020NVIDIASA,lu2021deepxde}, the ``PINNs-TF2" package utilizes static computational graphs via Accelerated Linear Algebra (XLA) and Just-In-Time (JIT) compilers \cite{50530}, a technique also adopted by others, to significantly enhance the speed of Physics-Informed Neural Networks (PINNs). This approach yields considerable improvements in training times over TensorFlow V1 \cite{Raissi2019PhysicsinformedNN}. Through our package, we have showcased its versatility by implementing nine distinct examples, achieving an impressive peak speed-up of 119.96x compared to previous implementations in TensorFlow V1. Also, with the incorporation of the Hydra framework \cite{Yadan2019Hydra}, ``PINNs-TF2'' refines user experience by distilling PDE problems into assorted samplers and boundary conditions.

Our results suggest that the exclusive use of the JIT compiler strikes an ideal equilibrium between speed-up and errors by obviating redundant graph constructions during gradient operations within TensorFlow's computational graph. Moreover, we elucidate the influence of batch sizes and the quantity of trainable parameters on the performance of implemented examples with our package. We hope our package serves as an indispensable tool for researchers delving into the PINNs domain.

\section{PINNs-TF2 Package}
In this section, we provide a brief summary of the problem setup and outline how our package works.

\begin{figure}
  \centering
  \includegraphics[width=\linewidth]{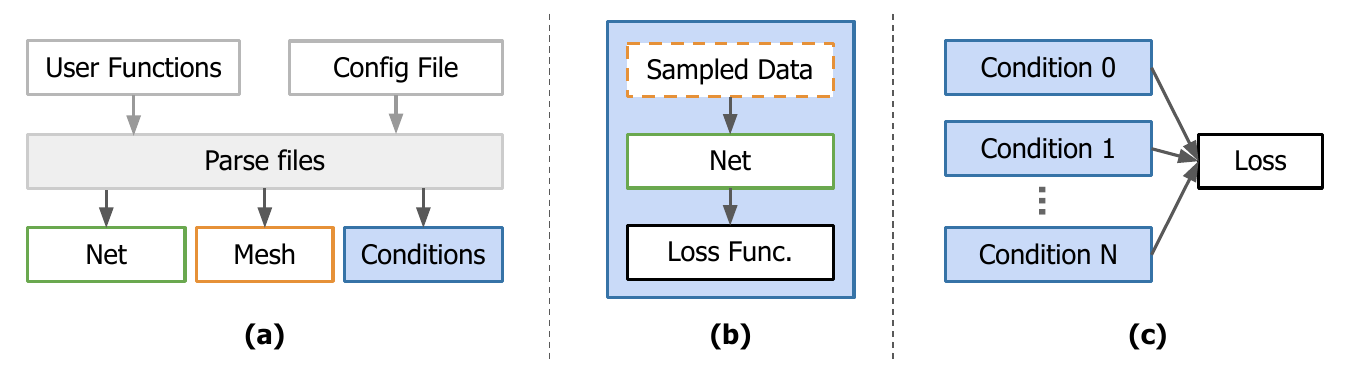}
  \caption{Overview of the simplified PINNs-TF2 framework: 
  \textbf{(a)} Users provide a config file and define reading data and PDE functions. The package then processes them to initialize a neural network, formulate a mesh for the data, and establish conditions.
  \textbf{(b)} Elucidates the components of the conditions, including data sampled from the mesh, shared neural networks, and designated loss functions.
  \textbf{(c)} During training, cumulative loss from each condition directs backpropagation. This process is compiled with the XLA compiler.}
  \label{fig:overview}
\end{figure}

\subsection{Problem Setup}
We adopt the problem definition from \cite{Raissi2019PhysicsinformedNN}. The study focuses on parametric and nonlinear PDEs with the structure:
$$
u_t + \mathcal{N}[u; \lambda],\quad x \in \Omega ,\quad t \in [0, T]
$$
Where $u(t, x)$ is the unobservable solution, $\mathcal{N}[.; \lambda]$ is a nonlinear operator influenced by the parameter $\lambda$, and $\Omega$ is a subset of ${\mathbb{R}}^{D}$. Two core issues are addressed: Data-driven solution (forward problem) \cite{Raissi2017NumericalGP, Raissi2016InferringSO} focuses on revealing the hidden state $u(t, x)$ for a given $\lambda$. Data-driven discovery (inverse problem) \cite{Raissi2017NumericalGP, Raissi2017HiddenPM, Rudy2016DatadrivenDO} seeks the optimal $\lambda$ values based on observed data. 

There are two algorithm approaches based on data types: continuous time and discrete time models. The former employs new spatio-temporal approximators, and the latter uses specific implicit Runge-Kutta methods. For further information, please refer to \cite{Raissi2019PhysicsinformedNN}.

\subsection{Implementation}

\paragraph{PINNs-TF2 Workflow.}
\label{sec22}
Our package streamlines the process of addressing both forward and inverse challenges in discrete and continuous contexts linked to nonlinear partial differential equations. Initially, it processes configuration files using Hydra \cite{Yadan2019Hydra} to retrieve specifications such as spatial/temporal ranges, the number of samples, boundary conditions, and neural network attributes like layer count. Subsequently, it loads the user-specified PDE function and a function for loading data. With this data at hand, the relevant conditions, mesh based on data, and a neural network are initialized. Each condition can have its unique loss function (e.g. periodic boundary conditions calculate loss from the difference between predicted and actual values at the periodic boundary). All conditions share a common neural network. Both the training and evaluation phases are compiled using the XLA compiler. An illustrative overview of this workflow can be seen in Figure \ref{fig:overview}. 
\vspace{-0.5em}
\paragraph{Compile with \texttt{tf.function}.}
When \texttt{tf.function} is set to \texttt{jit\_compile=False}, TensorFlow translates Python functions into a static computational graph, optimized through pattern-matching rewrites. This approach, however, doesn't generate new code and relies on a limited set of predefined kernels, which can sometimes restrict its flexibility and optimization potential. In our work, this mode for compiling training and evaluation steps is referred to as ``TF2''.

On the other hand, when \texttt{tf.function} employs \texttt{jit\_compile=True}, the XLA compiler \cite{50530}, leveraging Just-In-Time (JIT) compilation, converts TensorFlow's computation graphs into highly optimized machine code right before execution. JIT offers several advantages: by fusing multiple operations, operating on the High Level Optimiser Internal Representation (HLO IR), and tailoring the code for the nuances of the target hardware, it significantly enhances both speed and memory efficiency \cite{mlir,50530}. JIT-optimized processes reduce overhead in PINNs' repetitive gradient computations, with XLA's static graph enhancing computational efficiency. In our architecture, we maintain consistent input shapes to avoid XLA compiler overhead from shape changes \cite{Subramani2020EnablingFD}. For the purposes of our experiments, this mode is termed ``JIT''.

\vspace{-0.5em}
\paragraph{Mixed Precision.}
We also employ TensorFlow's Mixed Precision using float16, blending FP16 and FP32 to accelerate training and conserve memory on GPUs \cite{Micikevicius2017MixedPT}. We labeled this mode "AMP".

\section{Experiments}
\label{sec3}
In this section, we assess the performance of TF2, JIT, and AMP across 8 diverse examples, focusing on error maintenance and speed-up. We also demonstrate their efficacy with a large-scale dataset by implementing a real-world example.

\vspace{-0.7em}
\paragraph{Hardware Setup.}
All tests were carried out on a single NVIDIA Quadro RTX 8000 GPU to maintain uniformity and repeatability.
\vspace{-0.7em}
\paragraph{Speed-up Metric.}

We measured the median time for a single iteration in each case and compared it to the original TensorFlow V1 (TF1) implementations\footnote{For instances in section \ref{sec1}, we reference the code from \href{https://github.com/maziarraissi/PINNs}{\texttt{https://github.com/maziarraissi/PINNs}}, and for the instance in section \ref{sec2}, we consult the code from \href{https://github.com/maziarraissi/HFM}{\texttt{https://github.com/maziarraissi/HFM}}.}. The speed-up is measured by dividing the duration from the TF1 version by the duration of each specific scenario in TensorFlow V2.
\vspace{-0.7em}
\paragraph{Mean Relative Error Metric.} We calculate average relative errors for each example. Error nature may vary by problem; see Supplementary Materials Section \ref{error} for details.
\subsection{Evaluation of Various Acceleration Techniques}
\label{sec1}
We measure the efficacy of acceleration techniques across various examples, including the Continuous Forward Schrodinger, Discrete Forward Allen–Cahn (AC), Continuous Inverse Navier-Stokes (NS), and Discrete Inverse Korteweg-de Vries (KdV) Equations. We also explore the Burgers' Equation in all modes. For in-depth insights about examples, see the Supplementary Materials Section \ref{examples} and \cite{Raissi2019PhysicsinformedNN}. Our benchmarks compare JIT compiler and AMP combinations against a non-accelerated baseline.

\begin{figure}
  \centering
  \includegraphics[width=\linewidth]{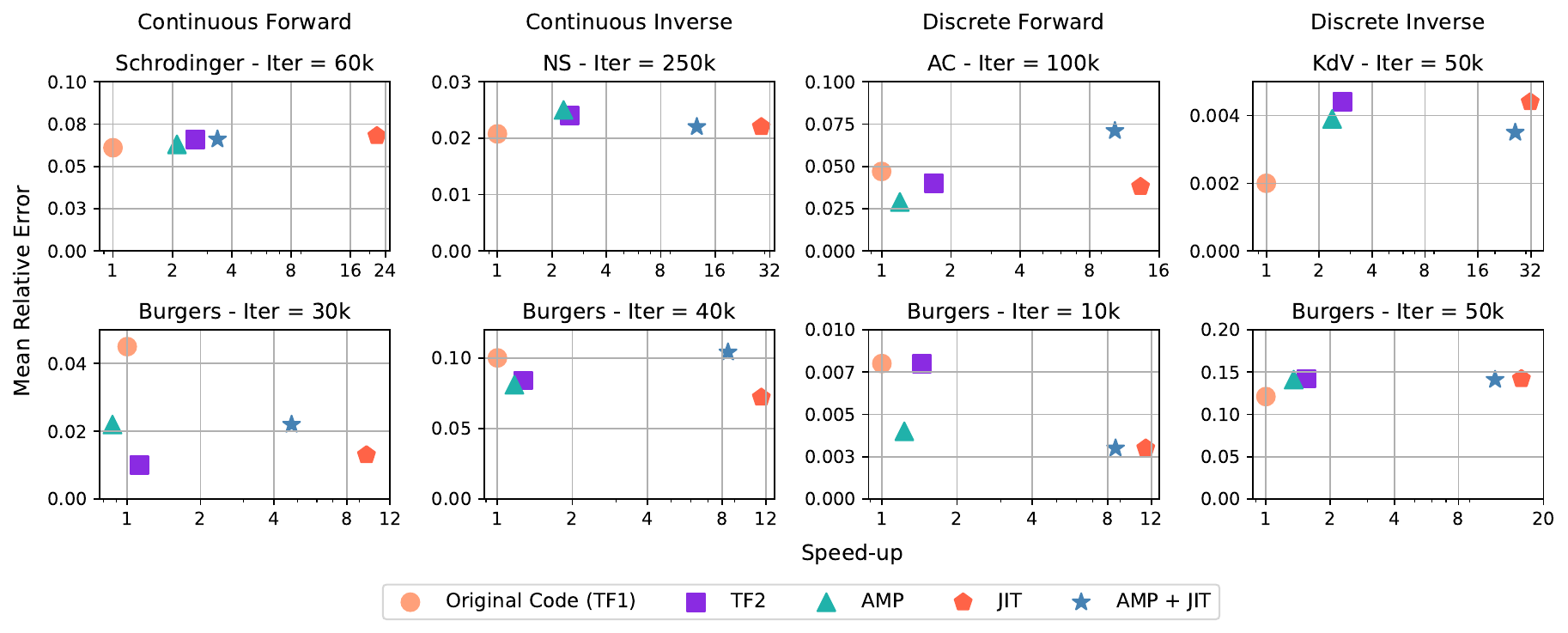}
  \caption{Each subplot denotes a unique problem, with its specific iteration count indicated at the top. The logarithmic x-axis shows speed-up relative to TF1, and the y-axis the mean error, highlighting JIT compiler boosts speed without added error.}
  \label{fig:perf}
\end{figure}

In Table \ref{tab:average-perf}, we present the average speed-ups of our examples compared to TensorFlow V1. By solely utilizing the JIT compiler, we achieved average speed-up of 18.12 without any compromise in accuracy. This advantage is visually represented in Figure \ref{fig:perf}, which plots both the speed-up and mean relative errors. The KdV example registered the highest speed-up, peaking at 31.75. Conversely, our performance did not benefit from using AMP, a limitation possibly due to hardware constraints or implementation overheads. Notably, TF2 outperformed TF1 with an average speed-up of 1.81. Moreover, our data points towards a decline in speed-up when AMP and JIT are combined, a phenomenon potentially resulting from the mixing precision and JIT overheads.

\begin{table*}
\caption{Average speed-ups for the eight examples from Section 3.1, benchmarked against TensorFlow V1 using different acceleration methods. Among the methods, the JIT compiler alone stands out as the most effective accelerator.\\}
\label{tab:average-perf}
\centering
\begin{tabular*}{\linewidth}{@{\extracolsep{\fill}}lcccccc}
    \toprule
    &
    TF2 & 
    JIT &
    AMP & 
    AMP+JIT 
    \\
    \toprule
    Avg. speed-up w.r.t. TF1 & 1.81 & \textbf{18.12} &  1.58 & 10.76 \\
    \bottomrule
\end{tabular*}
\end{table*}

\subsection{Assessing the Impact of Batch Size and Number Trainable Parameters}
\label{sec2}
We explore the influence of batch sizes and the count of trainable parameters on the efficiency of models harnessing our accelerators. This section focuses on the computational intricacies of modeling a three-dimensional physiological blood flow inside a genuine intracranial aneurysm (ICA) using the 3D Navier-Stokes equation. Given the dataset's vastness, encompassing 29 million data points spanning spatial, and temporal domains, and five solutions, we reshuffle it each epoch, sampling according to the batch size. For further insights, consult \cite{Raissi2018HiddenFM,Raissi2020HiddenFM}. 

\begin{figure}
  \centering
  \includegraphics[width=\linewidth]{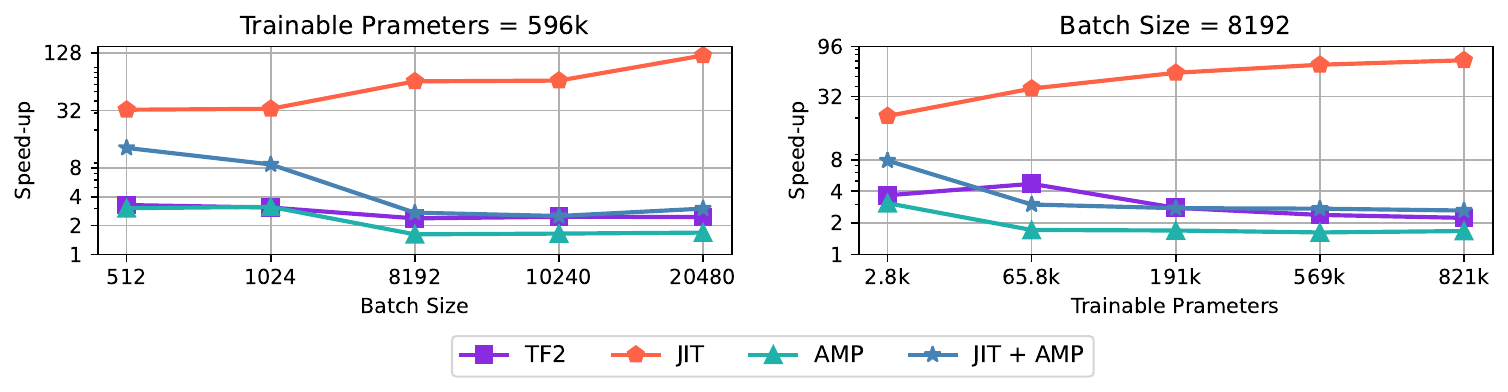}
  \caption{The left plot depicts the escalating efficiency benefits from the JIT compiler with increasing batch sizes. In contrast, the right plot reveals that adjusting the number of trainable parameters by altering the neural network's layer count enhances the JIT's effectiveness, while the speed-up in other configurations diminishes.}
  \label{fig:batch}
\end{figure}

We assess speed-up metrics across various settings. Initially, with all other attributes fixed, we solely vary the batch size for a model with 596k trainable parameters. The left plot of Figure \ref{fig:batch} reveals that the JIT's efficiency surges with larger batch sizes. At a batch size of 20480, the JIT's speed-up peaks at 119.96, significantly reducing training durations. This is likely due to the JIT compiler's capacity to optimize memory usage of computational graphs, such as through fusion, allowing more room for increased batch sizes. In our subsequent experiment, with a fixed batch size of 8192, we adjusted the number of trainable parameters by modifying the neural network's layer depth. While other configurations witnessed a performance decline, JIT's efficacy rose, as showcased in the right plot of Figure \ref{fig:batch}. The highest speed-up relative to TF1 reaches 70.99 with over 821k trainable parameters and a batch size of 8192. This section underscores the JIT compiler's advantage, especially for large batch sizes and a high count of trainable parameters.

\section{Conclusions}
In this package, we underscore the significance of compilers in TensorFlow, demonstrating their capability to boost performance over standard TensorFlow implementations. Through 9 varied examples, we illustrate the versatility of ``PINNs-TF2'' across diverse challenges. Especially for large batch sizes, the use of XLA and JIT compilers has yielded a remarkable 119x speed-up compared to TensorFlow V1. Interestingly, in our tests, mixed precision reduced the speed-up, suggesting that newer GPUs (i.g. NVIDIA A100) and the adoption of TensorFloat-32 might address this issue \cite{Choquette2021NVIDIAAT,Hennigh2020NVIDIASA}. We believe our package will be valuable to research across various domains.

\newpage
{
\small
\bibliographystyle{ieee_fullname}
\bibliography{main}

\begin{thebibliography}{10}\itemsep=-1pt

\bibitem{Chen2019PhysicsinformedNN}
Yuyao Chen, Lu Lu, George~Em Karniadakis, and Luca~Dal Negro.
\newblock Physics-informed neural networks for inverse problems in nano-optics and metamaterials.
\newblock {\em Optics express}, 28 8:11618--11633, 2019.

\bibitem{Choquette2021NVIDIAAT}
Jack Choquette, Wishwesh Gandhi, Olivier Giroux, Nick Stam, and Ronny Krashinsky.
\newblock Nvidia a100 tensor core gpu: Performance and innovation.
\newblock {\em IEEE Micro}, 41:29--35, 2021.

\bibitem{Haghighat2021APD}
Ehsan Haghighat, Maziar Raissi, Adrian Moure, H{\'e}ctor G{\'o}mez, and Ruben Juanes.
\newblock A physics-informed deep learning framework for inversion and surrogate modeling in solid mechanics.
\newblock {\em Computer Methods in Applied Mechanics and Engineering}, 379:113741, 2021.

\bibitem{Hennigh2020NVIDIASA}
Oliver Hennigh, Susheela Narasimhan, Mohammad~Amin Nabian, Akshay Subramaniam, Kaustubh~Mahesh Tangsali, Max Rietmann, Jos{\'e} del {\'A}guila~Ferrandis, Wonmin Byeon, Zhiwei Fang, and Sanjay Choudhry.
\newblock Nvidia simnet: an ai-accelerated multi-physics simulation framework.
\newblock In {\em International Conference on Conceptual Structures}, 2020.

\bibitem{iserles2009first}
Arieh Iserles.
\newblock {\em A first course in the numerical analysis of differential equations}.
\newblock Number~44. Cambridge university press, 2009.

\bibitem{Jeong2023ACP}
Hyogu Jeong, C.P. Batuwatta-Gamage, Jinshuai Bai, Yi~Min Xie, Charith Rathnayaka, Ying Zhou, and Yuantong Gu.
\newblock A complete physics-informed neural network-based framework for structural topology optimization.
\newblock {\em Computer Methods in Applied Mechanics and Engineering}, 2023.

\bibitem{mlir}
Chris Lattner, Mehdi Amini, Uday Bondhugula, Albert Cohen, Andy Davis, Jacques Pienaar, River Riddle, Tatiana Shpeisman, Nicolas Vasilache, and Oleksandr Zinenko.
\newblock {{MLIR}}: Scaling compiler infrastructure for domain specific computation.
\newblock In {\em 2021 {{IEEE/ACM}} International Symposium on Code Generation and Optimization (CGO)}, pages 2--14, 2021.

\bibitem{lu2021deepxde}
Lu Lu, Xuhui Meng, Zhiping Mao, and George~Em Karniadakis.
\newblock {DeepXDE}: A deep learning library for solving differential equations.
\newblock {\em SIAM Review}, 63(1):208--228, 2021.

\bibitem{Mao2020PhysicsinformedNN}
Zhiping Mao, Ameya~Dilip Jagtap, and George~Em Karniadakis.
\newblock Physics-informed neural networks for high-speed flows.
\newblock {\em Computer Methods in Applied Mechanics and Engineering}, 360:112789, 2020.

\bibitem{McClenny2021TensorDiffEqSM}
Levi~D. McClenny, Mulugeta~A. Haile, and Ulisses~M. Braga-Neto.
\newblock Tensordiffeq: Scalable multi-gpu forward and inverse solvers for physics informed neural networks.
\newblock {\em ArXiv}, abs/2103.16034, 2021.

\bibitem{Micikevicius2017MixedPT}
Paulius Micikevicius, Sharan Narang, Jonah Alben, Gregory~Frederick Diamos, Erich Elsen, David Garc{\'i}a, Boris Ginsburg, Michael Houston, Oleksii Kuchaiev, Ganesh Venkatesh, and Hao Wu.
\newblock Mixed precision training.
\newblock {\em ArXiv}, abs/1710.03740, 2017.

\bibitem{Paszke2017AutomaticDI}
Adam Paszke, Sam Gross, Soumith Chintala, Gregory Chanan, Edward Yang, Zach DeVito, Zeming Lin, Alban Desmaison, Luca Antiga, and Adam Lerer.
\newblock Automatic differentiation in pytorch.
\newblock 2017.

\bibitem{Paszke2019PyTorchAI}
Adam Paszke, Sam Gross, Francisco Massa, Adam Lerer, James Bradbury, Gregory Chanan, Trevor Killeen, Zeming Lin, Natalia Gimelshein, Luca Antiga, Alban Desmaison, Andreas K{\"o}pf, Edward Yang, Zach DeVito, Martin Raison, Alykhan Tejani, Sasank Chilamkurthy, Benoit Steiner, Lu Fang, Junjie Bai, and Soumith Chintala.
\newblock Pytorch: An imperative style, high-performance deep learning library.
\newblock In {\em Neural Information Processing Systems}, 2019.

\bibitem{Raissi2017HiddenPM}
Maziar Raissi and George~Em Karniadakis.
\newblock Hidden physics models: Machine learning of nonlinear partial differential equations.
\newblock {\em ArXiv}, abs/1708.00588, 2017.

\bibitem{Raissi2016InferringSO}
Maziar Raissi, Paris Perdikaris, and George~Em Karniadakis.
\newblock Inferring solutions of differential equations using noisy multi-fidelity data.
\newblock {\em J. Comput. Phys.}, 335:736--746, 2016.

\bibitem{Raissi2017NumericalGP}
Maziar Raissi, Paris Perdikaris, and George~Em Karniadakis.
\newblock Numerical gaussian processes for time-dependent and nonlinear partial differential equations.
\newblock {\em SIAM J. Sci. Comput.}, 40, 2017.

\bibitem{Raissi2019PhysicsinformedNN}
Maziar Raissi, Paris Perdikaris, and George~Em Karniadakis.
\newblock Physics-informed neural networks: A deep learning framework for solving forward and inverse problems involving nonlinear partial differential equations.
\newblock {\em J. Comput. Phys.}, 378:686--707, 2019.

\bibitem{Raissi2018HiddenFM}
Maziar Raissi, Alireza Yazdani, and George~Em Karniadakis.
\newblock Hidden fluid mechanics: A navier-stokes informed deep learning framework for assimilating flow visualization data.
\newblock {\em ArXiv}, abs/1808.04327, 2018.

\bibitem{Raissi2020HiddenFM}
Maziar Raissi, Alireza Yazdani, and George~Em Karniadakis.
\newblock Hidden fluid mechanics: Learning velocity and pressure fields from flow visualizations.
\newblock {\em Science}, 367:1026 -- 1030, 2020.

\bibitem{RashtBehesht2021PhysicsInformedNN}
Majid Rasht-Behesht, Christian Huber, Khemraj Shukla, and George~Em Karniadakis.
\newblock Physics‐informed neural networks (pinns) for wave propagation and full waveform inversions.
\newblock {\em Journal of Geophysical Research: Solid Earth}, 127, 2021.

\bibitem{Rudy2016DatadrivenDO}
Samuel~H. Rudy, Steven~L. Brunton, Joshua~L. Proctor, and J.~Nathan Kutz.
\newblock Data-driven discovery of partial differential equations.
\newblock {\em Science Advances}, 3, 2016.

\bibitem{50530}
Amit Sabne.
\newblock Xla : Compiling machine learning for peak performance, 2020.

\bibitem{Subramani2020EnablingFD}
Pranav Subramani, Nicholas Vadivelu, and Gautam Kamath.
\newblock Enabling fast differentially private sgd via just-in-time compilation and vectorization.
\newblock In {\em Neural Information Processing Systems}, 2020.

\bibitem{Yadan2019Hydra}
Omry Yadan.
\newblock Hydra - a framework for elegantly configuring complex applications.
\newblock Github, 2019.

\end{thebibliography}
}

\newpage
\pagebreak
\appendix

\section{Appendix}
This supplementary document expands on the primary paper in the following ways:
\begin{enumerate}
\item Additional information about the PINNs-TF2 Workflow (supplements \textbf{Section \ref{sec22}}).
\item Presents deeper insights into relative errors for evaluation and training for every problem (supplements \textbf{Section \ref{sec3}}).
\item Provides detailed conversations about the examples implemented using our package, including the associated errors and speed-ups (complements \textbf{Section \ref{sec3}}).
\end{enumerate}
\section{PINNs-Torch Workflow}
\begin{figure}
  \centering
  \includegraphics[width=\linewidth]{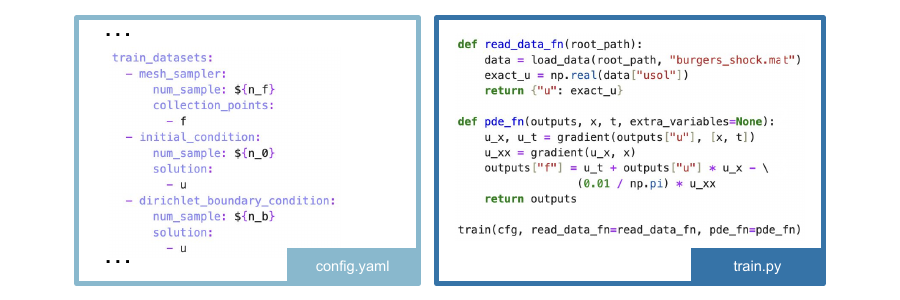}
  \caption{An example of a basic config file and custom functions for the continuous forward Burgers' equation. Users are required to set up a config file and function for the PINNs-TF2 package to process the data and determine the PDE.}
  \label{fig:overview-supp}
\end{figure}

Users should establish a config file interpreted by Hydra, which in turn triggers the relevant classes. They must also delineate functions for fetching data and defining the PDE. An illustration of this setup, using a file and custom functions, can be seen in Figure \ref{fig:overview-supp}. This package leverages these user definitions to solve the PDE.
\section{Errors}
\label{error}
\paragraph{Error Notations.}
We use \( \text{Err} \) as a unified symbol representing either mean squared error (MSE) or sum squared error (SSE). Specifically, \( \text{Err}_0 \) is the error in initial conditions, \( \text{Err}_b \) in boundary conditions, \( \text{Err}_c \) at collection points, \( \text{Err}_s \) in sampled solutions, and \( \text{Err}^i \) at time step \( i \).

\paragraph{Relative Errors.}
We measure errors between predicted and exact solutions using the \(\ell_2\) norm:
\begin{equation}
    \frac{\| u_{\text{pred}} - u_{\text{target}} \|_{2}}{\| u_{\text{target}} \|_{2}}
\end{equation}
For variables in the inverse problem, we use:
\begin{equation}
    \frac{| \lambda_{\text{pred}} - \lambda_{\text{target}} |}{| \lambda_{\text{target}} |}
\end{equation}

\section{Examples}
\label{examples}
In this section, we summarize examples from our main paper. For detailed insights on the first 8 examples, refer to \cite{Raissi2019PhysicsinformedNN} and for the 3D Navier-Stokes equation in section 3.2, see \cite{Raissi2018HiddenFM,Raissi2020HiddenFM}.

\paragraph{Continuous Forward Schrodinger Equation.}

\begin{table*}
\caption{The problem setup for continuous forward Schrodinger equation.\\}
\label{tab:schr-eq}
\centering
\begin{tabular*}{\linewidth}{@{\extracolsep{\fill}}lr}
    \toprule
    Continuous Forward Schrodinger Equation \\
    \toprule
    PDE equations & \( 
    \begin{aligned}
        f_u &= u_t + 0.5v_{xx} + v(u^2 +v^2), \\
    f_v &= v_t + 0.5u_{xx} + u(u^2 +v^2) 
    \end{aligned} 
    \) \\
    \midrule
    Initial condition  & \( 
    \begin{aligned}
        u(0, x) = 2 \text{sech}(x), \\
        v(0, x) = 2 \text{sech}(x)
    \end{aligned} 
    \) \\
    \midrule
    Periodic boundary conditions & \( 
    \begin{aligned}
        u(t,-5) &= u(t, 5), \\
        v(t,-5) &= v(t, 5), \\
    u_x(t,-5) &= u_x(t, 5), \\
    v_x(t,-5) &= v_x(t, 5) 
    \end{aligned} 
    \) \\
    \midrule   
    The output of net & \( 
    \begin{aligned}
        [u(t,x),v(t,x)]
    \end{aligned} 
    \) \\
    \midrule   
    Layers of net & \([2] + 4 \times [100] +[2]\) \\
    \midrule   
    Sample count from collection points & 
    \( 20000\) \\
    \midrule   
    Sample count from the initial condition & \(50\) \\
    \midrule   
    Sample count from boundary conditions & \(50\) \\
    \midrule
    Loss function & \(\text{MSE}_0 + \text{MSE}_b + \text{MSE}_c\) \\
    
    \bottomrule
\end{tabular*}
\end{table*}

For the nonlinear Schrodinger equation given by:
\begin{align*}
ih_t + 0.5h_{xx} + {|h|}^2h &= 0, \\
h(0,x) &= 2 \text{sech}(x),\\
h(t,-5) &= h(t, 5),\\
h_x(t,-5) &= h_x(t, 5),
\end{align*}
with $x\in [-5,5]$, $t\in [0,\pi/2]$, and $h(t, x)$ as the complex solution, we partition $h(t, x)$ into its real part $u$ and imaginary part $v$. Thus, our complex-valued neural network representation is $[u(t, x), v(t, x)]$. The setup is detailed in Table \ref{tab:schr-eq}.

Prediction discrepancies are gauged against the test data using the relative $\ell_2$-norm. Table \ref{tab:schrodinger} showcases errors for $h(t, x)$, $u(t, x)$, and $v(t, x)$, plus the average error as mentioned in the primary study.

\begin{table*}
\caption{Comparison of different methods in terms of individual errors, mean error, and speed-up factor for continuous forward Schrodinger equation after 60,000 iterations.\\}
\label{tab:schrodinger}
\centering
\begin{tabular*}{\linewidth}{@{\extracolsep{\fill}}lccccc}
    \toprule
    \multirow{2}{*}{Method} & 
    \multicolumn{3}{c}{Relative Errors} &
    \multirow{2}{*}{Mean Relative Error} & 
    \multirow{2}{*}{Speed-up} \\
    \cmidrule(r){2-4}
    & $h(t, x)$
    & $v(t, x)$
    & $u(t, x)$
    &&\\
    \toprule
    Original Code (TF1) & 0.017 & 0.104 &  0.064 & \textbf{0.061} & 1 \\
    TF2 & 0.024 & 0.106 &  0.068 & 0.066 & 2.62 \\
    AMP & 0.022 & 0.103 &  0.062 & 0.063 & 2.11 \\
    JIT & 0.024 & 0.110 &  0.069 & 0.068 & \textbf{22.90} \\
    JIT + AMP & 0.023 & 0.109 &  0.065 & 0.066 & 3.38 \\
    \bottomrule
\end{tabular*}
\end{table*}

\paragraph{Continuous Inverse Navier-Stokes Equation.}
Given the 2D nonlinear Navier-Stokes equation:
\begin{align*}
u_t + \lambda_1(uu_x + vu_y) &= -p_x + \lambda_2(u_{xx} + u_{yy}), \\
v_t + \lambda_1(uv_x + vv_y) &= -p_y + \lambda_2(v_{xx} + v_{yy}),
\end{align*}
where $u(t, x, y)$ and $v(t, x, y)$ are the x and y components of the velocity field, and $p(t, x, y)$ is the pressure, we seek the unknowns $\lambda = (\lambda_1, \lambda_2)$. When required, we integrate the constraints:
\begin{align}
0 &= u_x + v_y, \nonumber \\
u &= \psi_y, \nonumber \\
v &= -\psi_x, \label{navier}
\end{align}
We use a dual-output neural network to approximate $[\psi(t, x, y), p(t, x, y)]$, leading to a physics-informed neural network $[f(t, x, y), g(t, x, y)]$. The setup is detailed in Table \ref{tab:nav-eq}.

Prediction discrepancies are assessed against a test dataset. Table \ref{tab:navier} displays the relative $\ell_2$-norm errors for both velocity components and the relative errors for the $\lambda$ parameters, alongside the average error referenced in the main paper.

\begin{table*}
\caption{The problem setup for the continuous inverse Navier-Stokes equation.}
\label{tab:nav-eq}
\centering
\begin{tabular*}{\linewidth}{@{\extracolsep{\fill}}lr}
    \toprule
    Continuous Inverse Navier-Stokes Equation & \\
    \toprule
    PDE equations & \( 
    \begin{aligned}
        f &=  u_t + \lambda_1 (u u_x + v u_y)
        + p_x - \lambda_2  (u_{xx} + u_{yy}), \\
        g &= v_t + \lambda_1 (u v_x + v  v_y) + p_y - \lambda_2  (v_{xx} + v_{yy})
    \end{aligned} 
    \) \\
    \midrule
    Assumptions & \( 
    \begin{aligned}
        u &= \psi_y, \\
        v &= -\psi_x
    \end{aligned} 
    \) \\
    \midrule   
    The output of net& \( 
    \begin{aligned}
        [\psi(t, x, y), p(t, x, y)]
    \end{aligned} 
    \) \\
    \midrule   
    Layers of net & \([3] + 8 \times [20] +[2]\) \\
    \midrule   
    Sample count from collection points & 
    \( 5000^{*}\) \\
    \midrule   
    Sample count from solutions & 
    \( 5000^{*}\) \\
    \midrule
    Loss function & \(\text{SSE}_s  + \text{SSE}_c\) \\
    \bottomrule
\end{tabular*}

\smallskip
\textit{*Same points used for collocation and solutions.}
\end{table*}

\begin{table*}
\caption{Comparison of different methods in terms of individual errors, mean error, and speed-up factor for continuous inverse Navier-Stokes equation after 250,000 iterations.\\}
\label{tab:navier}
\centering
\begin{tabular*}{\linewidth}{@{\extracolsep{\fill}}lcccccc}
    \toprule
    \multirow{2}{*}{Method} & 
    \multicolumn{4}{c}{Relative Errors} &
    \multirow{2}{*}{Mean Relative Error} & 
    \multirow{2}{*}{Speed-up} \\
    \cmidrule(r){2-5}
    & $v(t, x)$
    & $u(t, x)$
    & $\lambda_1$
    & $\lambda_2$
    &&\\
    \toprule
    Original Code (TF1) & 0.018 & 0.009 &  0.002 & 0.054 & \textbf{0.021} & 1 \\
    TF2 & 0.021 & 0.023 & 0.001 & 0.051 & 0.024 & 2.50 \\
    AMP & 0.028 & 0.022 & 0.001 & 0.050 & 0.025 & 2.32 \\
    JIT & 0.019 & 0.021 & 0.001 & 0.045 & 0.022 & \textbf{28.77} \\
    JIT + AMP & 0.021 & 0.026 &  0.001 & 0.038 & 0.022 & 12.66 \\
    \bottomrule
\end{tabular*}
\end{table*}

\paragraph{Discrete Forward Allen-Cahn Equation.}
Given the non-linear AC equation:
\begin{align*}
u_t - 0.0001u_{xx} + 5u^3 - 5u &= 0, \\
u(0, x) &= x^2 \cos(\pi x), \\
u(t,-1) &= u(t, 1), \\
u_x(t,-1) &= u_x(t, 1),
\end{align*}
with $x \in [-1, 1]$ and $t \in [0, 1]$, we adopt Runge–Kutta methods with q stages as described in \cite{Raissi2019PhysicsinformedNN,iserles2009first}. The neural network output is:
\begin{align*}
[u^{n+c_1}(x),\dots, u^{n+c_q}(x), u^{n+1}(x)]
\end{align*}
where $u^{n+c_j}$ is data at time $t^n + c_j \Delta t$. The problem setup can be found in Table \ref{tab:ac-eq}. We extract data from the exact solution at $t_0 = 0.1$ aiming to predict the solution at $t_1 = 0.9$ using a single time-step of $\Delta t = 0.8$. Table \ref{tab:ac} shows $\ell_2$-norm errors for $u(x)$ at $t_1$.

\begin{table*}
\caption{The problem setup for discrete forward Allen-Cahn equation.\\}
\label{tab:ac-eq}
\centering
\begin{tabular*}{\linewidth}{@{\extracolsep{\fill}}lr}
    \toprule
    Discrete Forward AC Equation & \\
    \toprule
    PDE equations & \( 
    \begin{aligned}
       f^{n+c_j} = 5.0 u^{n+c_j} - 5.0 (u^{n+c_j})^3 + 0.0001 u^{n+c_j}_{xx}
    \end{aligned} 
    \) \\
    \midrule
    Periodic boundary conditions & \( 
    \begin{aligned}
        u(t,-1) &= u(t, 1), \\
        u_x(t,-1) &= u_x(t, 1) 
    \end{aligned} 
    \) \\
    \midrule   
    The output of net& \( 
    \begin{aligned}
        [u^{n+c_1}(x),\dots, u^{n+c_q}(x), u^{n+1}(x)]
    \end{aligned} 
    \) \\
     \midrule   
    Layers of net & \([1] + 4 \times [200] +[101]\) \\
    \midrule   
    The number of stages (q) & 100 \\
    \midrule   
    Sample count from collection points at \(t_0\) & 
    \( 200^{*}\) \\
    \midrule
     Sample count from solutions at \(t_0\) & 
    \( 200^{*}\) \\
    \midrule
    \(t_0 \rightarrow t_1\) & 
    \( 0.1 \rightarrow 0.9\) \\
    \midrule
    Loss function & \(\text{SSE}^{0}_s  + \text{SSE}^{0}_c + \text{SSE}^{1}_b\) \\
    \bottomrule
\end{tabular*}
\smallskip
\textit{*Same points used for collocation and solutions.}
\end{table*}

\paragraph{Discrete Inverse Korteweg–de Vries Equation.}
Given the non-linear KdV equation:
\begin{align*}
u_t + \lambda_1 uu_x + \lambda_2 u_{xxx} = 0,
\end{align*}
we use Runge–Kutta methods with q stages to identify parameters $\lambda = (\lambda_1, \lambda_2)$. The network outputs:
\begin{align*}
[u^{n+c_1}(x),\dots, u^{n+c_{q-1}}(x), u^{n+c_{q}}(x)]
\end{align*}
with $u^{n+c_j} = u (t^n+c_j\Delta t, x)$ as data at time $t^n + c_j\Delta t$. Data is sampled at $t^n = 0.2$ and $t^{n+1} = 0.8$. See Table\ref{tab:kdv-eq} for problem details and Table \ref{tab:kdv} for relative errors of $\lambda_1$ and $\lambda_2$.

\begin{table*}
\caption{The problem setup for discrete inverse Korteweg–de Vries equation.\\}
\label{tab:kdv-eq}
\centering
\begin{tabular*}{\linewidth}{@{\extracolsep{\fill}}lr}
    \toprule
    Discrete Inverse KdV Equation & \\
    \toprule
    PDE equations & \( 
    \begin{aligned}
       f^{n+c_j} = -\lambda_1 u^{n+c_j}u_x^{n+c_j} - \lambda_2  u^{n+c_j}_{xxx}
    \end{aligned} 
    \) \\
    \midrule   
    The output of net& \( 
    \begin{aligned}
        [u^{n+c_1}(x),\dots, u^{n+c_{q-1}}(x), u^{n+c_{q}}(x)]
    \end{aligned} 
    \) \\
     \midrule   
    Layers of net & \([1] + 3 \times [50] +[50]\) \\
    \midrule   
    The number of stages (q) & \(50\) \\
    \midrule   
    Sample count from solutions at \(t_0\) & 
    \( 199^{*}\) \\
    \midrule   
    Sample count from collection points at \(t_0\) & 
    \( 199^{*}\) \\
    \midrule   
    Sample count from solutions at \(t_1\) & 
    \( 201^{*}\) \\
    \midrule   
    Sample count from collection points at \(t_1\) & 
    \( 201^{*}\) \\
    \midrule
    \(t_0 \rightarrow t_1\) & 
    \( 0.2 \rightarrow 0.8\) \\
    \midrule
    Loss function & \(\text{SSE}^{0}_s  + \text{SSE}^{0}_c + \text{SSE}^{1}_s  + \text{SSE}^{1}_c\) \\
    \bottomrule
\end{tabular*}
\smallskip
\textit{*Same points used for collocation and solutions.}
\end{table*}

\begin{table}
\caption{Comparison of different methods in terms of individual errors, mean error, and speed-up factor for discrete forward Allen-Cahn equation after 100,000 iterations.\\}
\label{tab:ac}
\centering
\begin{tabular}{lccc}
    \toprule
    \multirow{2}{*}{Method} & 
    \multicolumn{1}{c}{Relative Error} &
    \multirow{2}{*}{Mean Relative Error} &
    \multirow{2}{*}{Speed-up} \\
    \cmidrule(r){2-2}
    & $u(t, x)$
    &&\\
    \toprule
    Original Code (TF1) & 0.047 & 0.047 & 1 \\
    TF2 & 0.040 & 0.040 & 1.68 \\
    AMP & 0.029 & \textbf{0.029} & 1.20 \\
    JIT & 0.038 & 0.038 & \textbf{13.31} \\
    JIT + AMP & 0.071 & 0.071 & 10.30 \\
    \bottomrule
\end{tabular}
\end{table}

\begin{table*}
\caption{Comparison of different methods in terms of individual errors, mean error, and speed-up factor for discrete inverse Korteweg–de Vries equation after 50,000 iterations.\\}
\label{tab:kdv}
\centering
\begin{tabular*}{\linewidth}{@{\extracolsep{\fill}}lcccc}
    \toprule
    \multirow{2}{*}{Method} & 
    \multicolumn{2}{c}{Relative Errors} &
    \multirow{2}{*}{Mean Relative Error} &
    \multirow{2}{*}{Speed-up} \\
    \cmidrule(r){2-3}
    & $\lambda_1$
    & $\lambda_2$
    &&\\
    \toprule
    Original Code (TF1) & 0.003 & 0.0005 & \textbf{0.002} & 1 \\
    TF2 & 0.002 & 0.007 & 0.004 & 2.72 \\
    AMP & 0.001 & 0.007 & 0.004 & 2.37 \\
    JIT & 0.002 & 0.007 & 0.004 & \textbf{31.75} \\
    JIT + AMP & 0.001 & 0.007 & 0.004 & 26.09 \\
    \bottomrule
\end{tabular*}
\end{table*}

\begin{table*}
\caption{The problem setup for continuous forward Burgers' equation.\\}
\label{tab:bcf-eq}
\centering
\begin{tabular*}{\linewidth}{@{\extracolsep{\fill}}lr}
    \toprule
    Continuous Forward Burgers' Equation & \\
    \toprule
    PDE equations & \( 
    \begin{aligned}
       f = u_t + uu_x - (0.01 /\pi) u_{xx}
    \end{aligned} 
    \) \\
     \midrule
    Initial conditions & \( 
    \begin{aligned}
        u(0, x) = -\sin(\pi x)
    \end{aligned} 
    \) \\
    \midrule
    Dirichlet boundary conditions & \( 
    \begin{aligned}
        u(t,-1) = u(t, 1) = 0
    \end{aligned} 
    \) \\
    \midrule   
    The output of net& \( 
    \begin{aligned}
        [u(t,x)]
    \end{aligned} 
    \) \\
    \midrule   
    Layers of net & \([2] + 8 \times [20] +[1]\) \\
    \midrule   
    Sample count from collection points & 
    \( 10000\) \\
    \midrule   
    Sample count from the initial condition & \(50\) \\
    \midrule   
    Sample count from boundary conditions & \(50\) \\
    \midrule
    Loss function & \(\text{MSE}_0 + \text{MSE}_b + \text{MSE}_c\) \\
    \bottomrule
\end{tabular*}
\end{table*}

\paragraph{Continuous Forward Burgers' Equation.}
Given the Burgers' equation:
\begin{align*}
u_t + uu_x - (0.01/\pi)u_{xx} = 0,
\end{align*}
with domain $x \in [-1, 1]$ and $t \in [0, 1]$, and the initial and boundary conditions:
\begin{align*}
u(0, x) &= -\sin(\pi x), \\
u(t,-1) &= 0, \\
u(t, 1) &= 0,
\end{align*}
we aim to determine the solution $u(t, x)$. Refer to Table \ref{tab:bcf-eq} for problem details and Table \ref{tab:bcf} for the relative error of $u(t, x)$.
\begin{table}
\caption{Comparison of different methods in terms of individual errors, mean error, and speed-up factor for continuous forward Burgers' equation after 30,000 iterations.\\}
\label{tab:bcf}
\centering
\begin{tabular}{lccc}
    \toprule
    \multirow{2}{*}{Method} & 
    \multicolumn{1}{c}{Relative Error} &
    \multirow{2}{*}{Mean Relative Error} &
    \multirow{2}{*}{Speed-up} \\
    \cmidrule(r){2-2}
    & $u(t, x)$
    &&\\
    \toprule
    Original Code (TF1) & 0.045 & 0.045 & 1 \\
    TF2 & 0.010 & \textbf{0.010} & 1.12 \\
    AMP & 0.022 & 0.022 & 0.87 \\
    JIT & 0.013 & 0.013 & \textbf{9.60} \\
    JIT + AMP & 0.022 & 0.022 & 4.73 \\
    \bottomrule
\end{tabular}
\end{table}

\begin{table*}
\caption{The problem setup for continuous inverse Burgers' equation.\\}
\label{tab:bci-eq}
\centering
\begin{tabular*}{\linewidth}{@{\extracolsep{\fill}}lr}
    \toprule
    Continuous Inverse Burgers' Equation & \\
    \toprule
    PDE equations & \( 
    \begin{aligned}
       f = u_t + \lambda_1 uu_x - \lambda_2 u_{xx}
    \end{aligned} 
    \) \\
     \midrule  
    The output of net& \( 
    \begin{aligned}
        [u(t,x)]
    \end{aligned} 
    \) \\
    \midrule   
    Layers of net & \([2] + 8 \times [20] +[1]\) \\
    \midrule   
    Sample count from collection points & 
    \( 2000^{*}\) \\
    \midrule   
    Sample count from solutions & \(2000^{*}\) \\
    \midrule
    Loss function & \(\text{MSE}_s + \text{MSE}_c\) \\
    \bottomrule
\end{tabular*}
\smallskip
\textit{*Same points used for collocation and solutions.}
\end{table*}

\paragraph{Continuous Inverse Burgers' Equation.}
Considering the equation:
\begin{align*}
u_t + \lambda_1uu_x - \lambda_2 u_{xx} = 0,
\end{align*}
we aim to both predict the solution $u(t, x)$ and determine the unknown parameters $\lambda = (\lambda_1, \lambda_2)$. For the problem configuration, see Table \ref{tab:bci-eq}. Relative errors for $u(t, x)$, $\lambda_1$, and $\lambda_2$ are in Table \ref{tab:bci}.

\begin{table*}
\caption{Comparison of different methods in terms of individual errors, mean error, and speed-up factor for continuous inverse Burgers' equation after 40,000 iterations.\\}
\label{tab:bci}
\centering
\begin{tabular*}{\linewidth}{@{\extracolsep{\fill}}lcccc}
    \toprule
    \multirow{2}{*}{Method} & 
    \multicolumn{2}{c}{Relative Errors} &
    \multirow{2}{*}{Mean Relative Error} &
    \multirow{2}{*}{Speed-up} \\
    \cmidrule(r){2-3}
    & $\lambda_1$
    & $\lambda_2$
    &&\\
    \toprule
    Original Code (TF1) & 0.003 & 0.196 & 0.100 & 1 \\
    TF2 & 0.009& 0.158 & 0.084 & 1.27 \\
    AMP & 0.006 & 0.155 & 0.081 & 1.17 \\
    JIT & 0.003 & 0.141 & \textbf{0.072} & \textbf{11.49} \\
    JIT + AMP & 0.040 & 0.167 & 0.104 & 8.44 \\
    \bottomrule
\end{tabular*}
\end{table*}

\paragraph{Discrete Forward Burgers' Equation.}
For this problem, we use data from $t_1 = 0.1$ to predict solutions at $t_2 = 0.9$ utilizing Runge-Kutta methods with q stages. The equation is:
\begin{align*}
f^{n+c_j} = u^{n+c_j}u_x^{n+c_j} - (0.01/\pi)u_{xx}^{n+c_j}
\end{align*}
Here, $u^n$ indicates information at time $t^n$. For more details, consult Table \ref{tab:bdf-eq} for the setup and Table \ref{tab:bdf} for relative errors of $u(t, x)$.

\begin{table}
\caption{Comparison of different methods in terms of individual errors, mean error, and speed-up factor for discrete forward Burgers' equation after 10,000 iterations.\\}
\label{tab:bdf}
\centering
\begin{tabular}{lccc}
    \toprule
    \multirow{2}{*}{Method} & 
    \multicolumn{1}{c}{Relative Error} &
    \multirow{2}{*}{Mean Relative Error} &
    \multirow{2}{*}{Speed-up} \\
    \cmidrule(r){2-2}
    & $u(t, x)$
    &&\\
    \toprule
    Original Code (TF1) & 0.008 & 0.008 & 1 \\
    TF2 & 0.008 & 0.008 & 1.45 \\
    AMP & 0.004 & 0.004 & 1.23 \\
    JIT & 0.003 & \textbf{0.003} & \textbf{11.39} \\
    JIT + AMP & 0.003 & \textbf{0.003} & 8.62 \\
    \bottomrule
\end{tabular}
\end{table}

\begin{table*}
\caption{The problem setup for discrete forward Burgers' equation.\\}
\label{tab:bdf-eq}
\centering
\begin{tabular*}{\linewidth}{@{\extracolsep{\fill}}lr}
    \toprule
    Discrete Forward Burgers' Equation & \\
    \toprule
    PDE equations & \( 
    \begin{aligned}
       f^{n+c_j} = u^{n+c_j}u_x^{n+c_j} - (0.01/\pi)u_{xx}^{n+c_j}
    \end{aligned} 
    \) \\
    \midrule
    Dirichlet boundary conditions & \( 
    \begin{aligned}
        u(t,-1) = u(t, 1) = 0
    \end{aligned} 
    \) \\
    \midrule   
    The output of net& \( 
    \begin{aligned}
        [u^{n+c_1}(x),\dots, u^{n+c_q}(x), u^{n+1}(x)]
    \end{aligned} 
    \) \\
         \midrule   
    Layers of net & \([1] + 3 \times [50] +[501]\) \\
    \midrule   
    The number of stages (q) & 500 \\
    \midrule   
    Sample count from collection points at \(t_0\) & 
    \( 250^{*}\) \\
    \midrule   
    Sample count from solutions at \(t_0\) & 
    \( 250^{*}\) \\
    \midrule
    \(t_0 \rightarrow t_1\) & 
    \( 0.1 \rightarrow 0.9\) \\
    \midrule
    Loss function & \(\text{SSE}^{0}_s  + \text{SSE}^{0}_c + \text{SSE}^{1}_b\) \\
    \bottomrule
\end{tabular*}
\smallskip
\textit{*Same points used for collocation and solutions.}
\end{table*}

\paragraph{Discrete Inverse Burgers' Equation.}
Similar to its forward counterpart, we utilize Runge-Kutta methods with q stages. The equation here is given by:
\begin{align*}
 f^{n+c_j} = \lambda_1 u^{n+c_j}u^{n+c_j}_x - \lambda_2 u^{n+c_j}_{xx}
\end{align*}
The goal is to determine $\lambda_1$ and $\lambda_2$. Data points are taken from $t=0.1$ to $t=0.9$. For more details, see Table \ref{tab:bdi-eq} for the problem setup and Table \ref{tab:bdi} for relative errors.

\paragraph{Continuous Forward 3D Navier-Stokes Equation.}
In this example, the fluid's dynamics are represented by the non-dimensional Navier-Stokes and continuity equations:
\begin{align*}
c_t + u c_x + v c_y + w c_z &= \text{\footnotesize Pec}^{-1} (c_{xx} + c_{yy} + c_{zz}),\\
u_t + u u_x + v u_y + w u_z &= - p_x + \text{\footnotesize Re}^{-1} (u_{xx} + u_{yy} + u_{zz}),\\
v_t + u v_x + v v_y + w v_z&= - p_y + \text{\footnotesize Re}^{-1} (v_{xx} + v_{yy} + v_{zz}),\\
w_t + u w_x + v w_y + w w_z &= - p_z + \text{\footnotesize Re}^{-1} (w_{xx} + w_{yy} + w_{zz}),\\
u_x + v_y + w_z &= 0.
\end{align*}

Velocity components are given by $\bold{u}=(u,v,w)$, and $p$ is the pressure. For the problem setup, refer to Table \ref{tab:3d-eq}. Adjustments were made in batch sizes, and hidden layers for parameter training.

\begin{table*}
\caption{The problem setup for continuous forward 3D Navier Stokes equation.\\}
\label{tab:3d-eq}
\centering
\begin{tabular*}{\linewidth}{@{\extracolsep{\fill}}lr}
    \toprule
    Continuous Forward 3D NS \\
    \toprule
    PDE equations & \( 
    \begin{aligned}
        e_1 =& c_t + (u c_x + v c_y + w c_z) \\&- (1.0 / \text{Pec})  (c_xx + c_yy + c_zz) \\
e_2 =& u_t + (u u_x + v u_y + w u_z) + p_x \\&- (1.0 / \text{Re}) (u_xx + u_yy + u_zz) \\
e_3 =& v_t + (u v_x + v v_y + w v_z) + p_y \\&- (1.0 / \text{Re}) (v_xx + v_yy + v_zz) \\
e_4 =& w_t + (u w_x + v w_y + w w_z) + p_z \\&- (1.0 / \text{Re}) (w_xx + w_yy + w_zz) \\
e_5 =& u_x + v_y + w_z
    \end{aligned} 
    \) \\
    \midrule   
    The output of net & \( 
    \begin{aligned}
        [c(t, x, y, z),u(t, x, y, z),v(t, x, y, z), \\ w(t, x, y, z),p(t, x, y, z)]
    \end{aligned} 
    \) \\
    \midrule   
    Layers of net & \([4] + 10 \times [250] +[5]\) \\
     \midrule   
    Batch size of collection points &
    \( 10000\) \\
    \midrule   
    Batch size of solutions in \(c(t, x, y, z)\)& 
    \( 10000\) \\
    \midrule
    
    Loss function & \(\text{MSE}_s + \text{MSE}_c\) \\
    
    \bottomrule
\end{tabular*}
\end{table*}

\begin{table*}
\caption{Comparison of different methods in terms of individual errors, mean error, and speed-up factor for discrete inverse Burgers' equation after 50,000 iterations.\\}
\label{tab:bdi}
\centering
\begin{tabular*}{\linewidth}{@{\extracolsep{\fill}}lcccc}
    \toprule
    \multirow{2}{*}{Method} & 
    \multicolumn{2}{c}{Error} &
    \multirow{2}{*}{Mean Error} &
    \multirow{2}{*}{Speed-up} \\
    \cmidrule(r){2-3}
    & $\lambda_1$
    & $\lambda_2$
    &&\\
    \toprule
    Original Code (TF1) & 0.003&0.239&\textbf{0.121} & 1 \\
    TF2 & 0.004 & 0.280 & 0.142 & 1.55 \\
    AMP & 0.004 & 0.278 & 0.141 & 1.35 \\
    JIT & 0.004 & 0.280 & 0.142 & \textbf{15.77} \\
    JIT + AMP & 0.004 & 0.278 & 0.141 & 11.87 \\
    \bottomrule
\end{tabular*}
\end{table*}

\begin{table*}
\caption{The problem setup for discrete inverse Burgers' equation.\\}
\label{tab:bdi-eq}
\centering
\begin{tabular*}{\linewidth}{@{\extracolsep{\fill}}lr}
    \toprule
    Discrete Inverse Burgers' Equation & \\
    \toprule
    PDE equations & \( 
    \begin{aligned}
       f^{n+c_j} = \lambda_1 u^{n+c_j}u^{n+c_j}_x - \lambda_2 u^{n+c_j}_{xx}
    \end{aligned} 
    \) \\
    \midrule   
    The output of net& \( 
    \begin{aligned}
        [u^{n+c_1}(x),\dots, u^{n+c_{q-1}}(x), u^{n+c_{q}}(x)]
    \end{aligned} 
    \) \\
         \midrule   
    Layers of net & \([1] + 4 \times [50] +[81]\) \\
    \midrule   
    The number of stages (q) & 81 \\
    \midrule   
    Sample count from collection points at \(t_0\) & 
    \( 199^{*}\) \\
    \midrule   
    Sample count from solutions at \(t_0\) & 
    \( 199^{*}\) \\
    \midrule
    Sample count from collection points at \(t_1\) & 
    \( 201^{*}\) \\
    \midrule   
    Sample count from solutions at \(t_1\) & 
    \( 201^{*}\) \\
    \midrule
    \(t_0 \rightarrow t_1\) & 
    \( 0.1 \rightarrow 0.9\) \\
    \midrule
    Loss function & \(\text{SSE}^{0}_s  + \text{SSE}^{0}_c + \text{SSE}^{1}_s + \text{SSE}^{1}_c\) \\
    \bottomrule
\end{tabular*}
\smallskip
\textit{*Same points used for collocation and solutions at each time step.}
\end{table*}

\end{document}